\documentclass[]{article}
\usepackage{graphics,graphicx,epsfig,rotating,color,ifthen}
\newcommand{\jpsi}{J/$\psi$}
\newcommand{\psip}{$\psi^\prime$}

\newcommand{\et}{$E_{\rm T}$}
\newcommand{\etsext}{$E_{\rm T}^i$}
\newcommand{\pt}{$p_{\rm T}$}

\newcommand{\npart}{$N_{\rm part}$}

\newcommand{\etalab}{$\eta_{\rm lab}$}
\newcommand{\lsim}{~\rlap{$<$}{\lower 1.0ex\hbox{$\sim$}}}
\newcommand{\rsim}{~\rlap{$>$}{\lower 1.0ex\hbox{$\sim$}}}
\newcommand{\ccbar}{$c\bar{c}$}
\newcommand{\ddbar}{$D\bar{D}$}
\newcommand{\massrange}{\mbox{$2.9<M<3.3$ GeV/c$^2$}}

\begin{document}

\centering{\bf \Large \jpsi\ azimuthal anisotropy relative to the reaction plane 
in Pb-Pb collisions at 158~GeV per nucleon} 

\vskip 0.5 cm 
F. Prino$^{10}$ on behalf of the NA50 collaboration

\medskip
\begin{small}
B.~Alessandro$^{10}$,
C.~Alexa$^{3}$,
R.~Arnaldi$^{10}$,
M.~Atayan$^{12}$,
S.~Beol\`e$^{10}$,
V.~Boldea$^{3}$,
P.~Bordalo$^{6}$,
G.~Borges$^{6}$,
C.~Castanier$^{2}$,
J.~Castor$^{2}$,
B.~Chaurand$^{9}$,
B.~Cheynis$^{11}$,
E.~Chiavassa$^{10}$,
C.~Cical\`o$^{4}$,
M.P.~Comets$^{8}$,
S.~Constantinescu$^{3}$,
P.~Cortese$^{1}$,
A.~De~Falco$^{4}$,
N.~De~Marco$^{10}$,
G.~Dellacasa$^{1}$,
A.~Devaux$^{2}$,
S.~Dita$^{3}$,
J.~Fargeix$^{2}$,
P.~Force$^{2}$,
M.~Gallio$^{10}$,
C.~Gerschel$^{8}$,
P.~Giubellino$^{10}$,
M.B.~Golubeva$^{7}$,
A.A.~Grigoryan$^{12}$,
S.~Grigoryan$^{12}$,
F.F.~Guber$^{7}$,
A.~Guichard$^{11}$,
H.~Gulkanyan$^{12}$,
M.~Idzik$^{10}$,
D.~Jouan$^{8}$,
T.L.~Karavicheva$^{7}$,
L.~Kluberg$^{9,5}$,
A.B.~Kurepin$^{7}$,
Y.~Le~Bornec$^{8}$,
C.~Louren\c co$^{5}$,
M.~Mac~Cormick$^{8}$,
A.~Marzari-Chiesa$^{10}$,
M.~Masera$^{10}$,
A.~Masoni$^{4}$,
M.~Monteno$^{10}$,
A.~Musso$^{10}$,
P.~Petiau$^{9}$,
A.~Piccotti$^{10}$,
J.R.~Pizzi$^{11}$,
F.~Prino$^{10}$,
G.~Puddu$^{4}$,
C.~Quintans$^{6}$,
L.~Ramello$^{1}$,
S.~Ramos$^{6}$,
L.~Riccati$^{10}$,
H.~Santos$^{6}$,
P.~Saturnini$^{2}$,
E.~Scomparin$^{10}$,
S.~Serci$^{4}$,
R.~Shahoyan$^{6}$,
F.~Sigaudo$^{10}$,
M.~Sitta$^{1}$,
P.~Sonderegger$^{5}$,
X.~Tarrago$^{8}$,
N.S.~Topilskaya$^{7}$,
G.L.~Usai$^{4}$,
E.~Vercellin$^{10}$,
L.~Villatte$^{8}$,
N.~Willis$^{8}$,
T.~Wu$^{8}$
\\
$^{~1}$ Universit\`a del Piemonte Orientale, Alessandria and
INFN-Torino, Italy;
$^{~2}$ LPC, Univ. Blaise Pascal and CNRS-IN2P3, Aubi\`ere, France;
$^{~3}$ IFA, Bucharest, Romania;
$^{~4}$ Universit\`a di Cagliari/INFN, Cagliari, Italy;
$^{~5}$ CERN, Geneva, Switzerland;
$^{~6}$ LIP, Lisbon, Portugal;
$^{~7}$ INR, Moscow, Russia;
$^{~8}$ IPN, Univ. de Paris-Sud and CNRS-IN2P3, Orsay, France;
$^{~9}$ Laboratoire Leprince-Ringuet,  Ecole Polytechnique and
CNRS-IN2P3,  Palaiseau,  France;
$^{10}$ Universit\`a di Torino/INFN, Torino, Italy;
$^{11}$ IPN, Univ. Claude Bernard Lyon-I and CNRS-IN2P3, Villeurbanne,
France;
$^{12}$ YerPhI, Yerevan, Armenia.
$^{13}$ Faculty of Physics and Applied Computer Science, AGH Univ., Cracow, Poland.
\end{small}

\abstract{
The J/$\psi$ azimuthal distribution relative to the reaction plane has been
measured by the NA50 experiment in Pb-Pb collisions at 158 GeV/nucleon.
Various physical mechanisms related to charmonium dissociation in the medium
created in the heavy ion collision are expected to introduce an anisotropy
in the azimuthal distribution of the observed J/$\psi$ mesons at SPS energies.
Hence, the measurement of J/$\psi$ elliptic anisotropy, quantified
by the Fourier coefficient v$_2$ of the J/$\psi$ azimuthal distribution
relative to the reaction plane, is an important tool to constrain
theoretical models aimed at explaining the anomalous J/$\psi$
suppression observed in Pb-Pb collisions.
We present the measured J/$\psi$ yields in different bins of
azimuthal angle relative to the reaction plane, as well as the resulting
values of the Fourier coefficient v$_{2}$ as a function of the
collision centrality and of the J/$\psi$ transverse momentum.
The reaction plane has been estimated from the azimuthal distribution of the
neutral transverse energy detected in an electromagnetic calorimeter.
The analysis has been performed on a data sample of about
100\,000 events, distributed in five centrality or p$_{\rm T}$ sub-samples.
The extracted v$_{2}$ values are significantly larger than zero
for non-central collisions and are seen to increase with p$_{\rm T}$.
}

\section{Introduction}

Charmonium production and suppression is one of the most powerful 
probes for a phase transition to deconfined matter in heavy-ion collisions 
at the energies of the CERN SPS.
In particular, the \jpsi\ suppression in proton-nucleus and 
nucleus-nucleus reactions has been intensively studied in the last 2 decades 
by the NA38 and NA50 experiments. 
The \jpsi\ suppression observed in proton-nucleus reactions is understood 
as due to absorption of charmonium states on ordinary nuclear matter with 
$\sigma_{abs}= 4.2\pm0.5$ mb ~\cite{Ale06}.
The \jpsi/Drell-Yan ratio measured in S-U and peripheral Pb-Pb collisions 
results to be in agreement with the expectation from ordinary nuclear 
absorption as measured in p-A reactions, while an anomalous extra 
suppression is present in semi-central and central Pb-Pb 
collisions~\cite{Ale05}.

Additional insight into charmonium suppression mechanisms can be obtained 
from the anisotropy of the overlap region of the projectile and target
nuclei in collisions with impact parameter 
b$>0$~\cite{Heisel,Wang,Zhu}.
The initial geometrical anisotropy gives rise to an observable  
anisotropy in particle distributions if the created system is 
interacting strongly enough to thermalize at an early stage 
and develop collective motion (flow). 
Hence, anisotropic transverse flow should be observed for \jpsi's formed by 
$c$-$\bar{c}$ recombination 
if the charm quarks have undergone strong enough re-scatterings 
leading them to thermalize in the first stages of the system evolution.
\jpsi\ flow is however not expected to be established at SPS energies
where early charm thermalization is unlikely and \ccbar\ 
recombination is negligible.
Nevertheless, other mechanisms related to \ccbar\ absorption in the medium 
created in the collision, essentially \ccbar\ dissociation by the
hard gluons present in the deconfined phase~\cite{Wang,Zhu}
and by co-moving hadrons~\cite{Heisel} are predicted as possible 
sources of \jpsi\ anisotropy already at SPS energies.

The azimuthal anisotropy is usually quantified from the coefficients 
of the Fourier series describing the particle azimuthal distribution:
\begin{equation}
\frac{dN}{d\varphi} \propto 1+\sum_{n=1}^{\infty}2 v_n \cos[n(\varphi-\Psi_{RP})]
\label{eq:Fourier}
\end{equation}
where $\Psi_{RP}$ is the reaction plane angle defined by the impact parameter 
vector in the transverse plane.
An azimuthal dependent \jpsi\ absorption pattern determined by 
the anisotropic geometrical shape of the nuclear overlap region 
is expected to give rise to a measurable second harmonic 
coefficient $v_2$, which describes an elliptic anisotropy.
It is anticipated~\cite{Wang,Zhu} that elliptic anisotropy due to \jpsi\ 
dissociation by gluons resulting from the formation of a deconfined medium
should vanish for 
peripheral collisions (where the critical temperature is not attained)
and for central collisions (because of the isotropic geometry of the 
overlap region), showing a sudden onset in correspondence of the phase 
transition and a maximum for semi-central collisions 
(\et$\approx$70-80 GeV in~\cite{Wang} or \npart$\approx$200-220 in~\cite{Zhu}).

\section{Experimental setup, data selection and reaction plane estimation}

The NA50 apparatus consists of a muon spectrometer
equipped with three detectors to measure centrality-related observables
on an event-by-event basis and specific devices for beam 
tagging and interaction vertex identification.
A detailed description of the detectors can be found in~\cite{Abr97a}.
Minimal details are given here for specific detectors relevant 
for the present analysis.

Anisotropy studies have been done on the sample of about 100,000 \jpsi's 
collected
by experiment NA50 in year 2000 with the SPS Pb beam at 
158~GeV/nucleon (see~\cite{Ale05}).
The \jpsi\ is detected via its $\mu^+\mu^-$ decay in the pseudo-rapidity
range $2.7\leq$ \etalab $\leq$ 3.9.
The analysis is performed in the dimuon kinematic domain 
\mbox{$0<y_{\rm cm}<1$} 
and \mbox{$-0.5 < \cos(\theta_{\rm CS}) < 0.5$}, 
where $y_{\rm cm}$ is the rapidity in the center-of-mass system and 
$\theta_{\rm CS}$ is the polar decay angle of the muons in the Collins-Soper 
reference frame.  

\begin{figure}[tb!]
\centering
\resizebox{0.45\textwidth}{!}
{\includegraphics*[bb= 68 151 506 519]{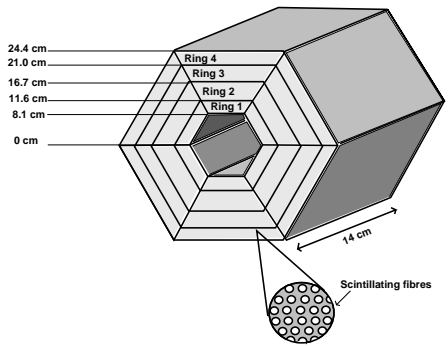}}
\caption{Geometrical segmentation of the electromagnetic calorimeter.}
\label{fig:emc}
\end{figure}

The study of the centrality dependence of the \jpsi\ anisotropy is based on
5 bins of the neutral transverse energy (\et) as measured by an
electromagnetic calorimeter (see fig.~\ref{fig:emc}).
This calorimeter is made up of lead and scintillating 
fibers and it measures event-by-event the transverse energy  
carried by neutral particles produced in the interaction 
(mostly due to $\pi^0 \rightarrow \gamma \gamma$ and to direct $\gamma$)
in the pseudo-rapidity window $1.1\leq$ \etalab $\leq 2.3$. 
The \et\ limits for the 5 bins are reported in table~\ref{tab:etbins}
together with the average values of impact parameter and number of participants
estimated by means of a Glauber calculation including the resolution of the
calorimeter.

\begin{table}[t!]
\begin{center}
\begin{tabular}{c| c c | c | c }
\hline\noalign{\smallskip}
Bin & E$_{\rm T,min}$ &  E$_{\rm T,max}$ & $\langle$b$\rangle$ 
& $\langle$N$_{\rm part}$$\rangle$\\
& (GeV) & (GeV) & (fm) & \\
\noalign{\smallskip}\hline\noalign{\smallskip}
1 & 10 & 30 & 10.5 &  64\\
2 & 30 & 50 & 8.3  & 134\\
3 & 50 & 70 & 6.5  & 203\\
4 & 70 & 90 & 4.8  & 271\\
5 & 90 & 120 & 2.7 & 342\\
\noalign{\smallskip}\hline
\end{tabular}
\caption{\et\ limits, average values of impact parameter and number of participants for the centrality bins used in this analysis.} 
\label{tab:etbins}
\end{center}
\end{table}

The reaction plane is estimated making use of the azimuthal segmentation
in six azimuthal sectors (sextants) of the electromagnetic calorimeter,
as represented in fig~\ref{fig:emc} where it can also be seen that
each sextant is further subdivided into four radial rings, each of 
them covering a different pseudo-rapidity range. 
The event plane angle $\Psi_n$ 
(estimator of the unknown reaction plane angle $\Psi_{RP}$) is given by:
\begin{equation}
\Psi_n=\frac{1}{n} \mathrm{tan}^{-1} \left[
\frac{ \sum_{i=1}^{6} ~ w^i ~ E_{\rm T}^i ~ \mathrm{sin}(n\Phi_i)}
{ \sum_{i=1}^{6} ~ w^i ~ E_{\rm T}^i ~ \mathrm{cos}(n\Phi_i)}
\right]
\label{eq:evpl}
\end{equation}
where $n$ is the considered Fourier harmonic, \etsext\ the neutral 
transverse energy measured in sextant $i$, and $\Phi_i$ the azimuthal 
angle defined by the center of sextant $i$ (see fig.~\ref{fig:emc}).
The weighting coefficients $w^i$ are introduced to make the event plane 
distribution isotropic and are defined as 
$<E_{\rm T}^{tot}>/(6 <E_{\rm T}^i>)$.
Their values range between 
0.994 and 1.012, resulting in a very small event-plane flattening 
correction.
The event plane $\Psi_2$ has been used to calculate the elliptic anisotropy 
\footnote{The event plane $\Psi_1$ could also be used,
but the resolution is worse~\cite{PoskVol}.}.
It is computed from the $\pi^\circ$ azimuthal distribution in the backward 
rapidity region, where, at SPS energies, pions show positive 
$v_2$~\cite{NA49flow,WA98flow,NA50flow} and therefore it
is directed in-plane (i.e. parallel to the reaction plane).

\begin{figure}[ht!]
\centering
\resizebox{0.31\textwidth}{!}{%
\includegraphics*[]{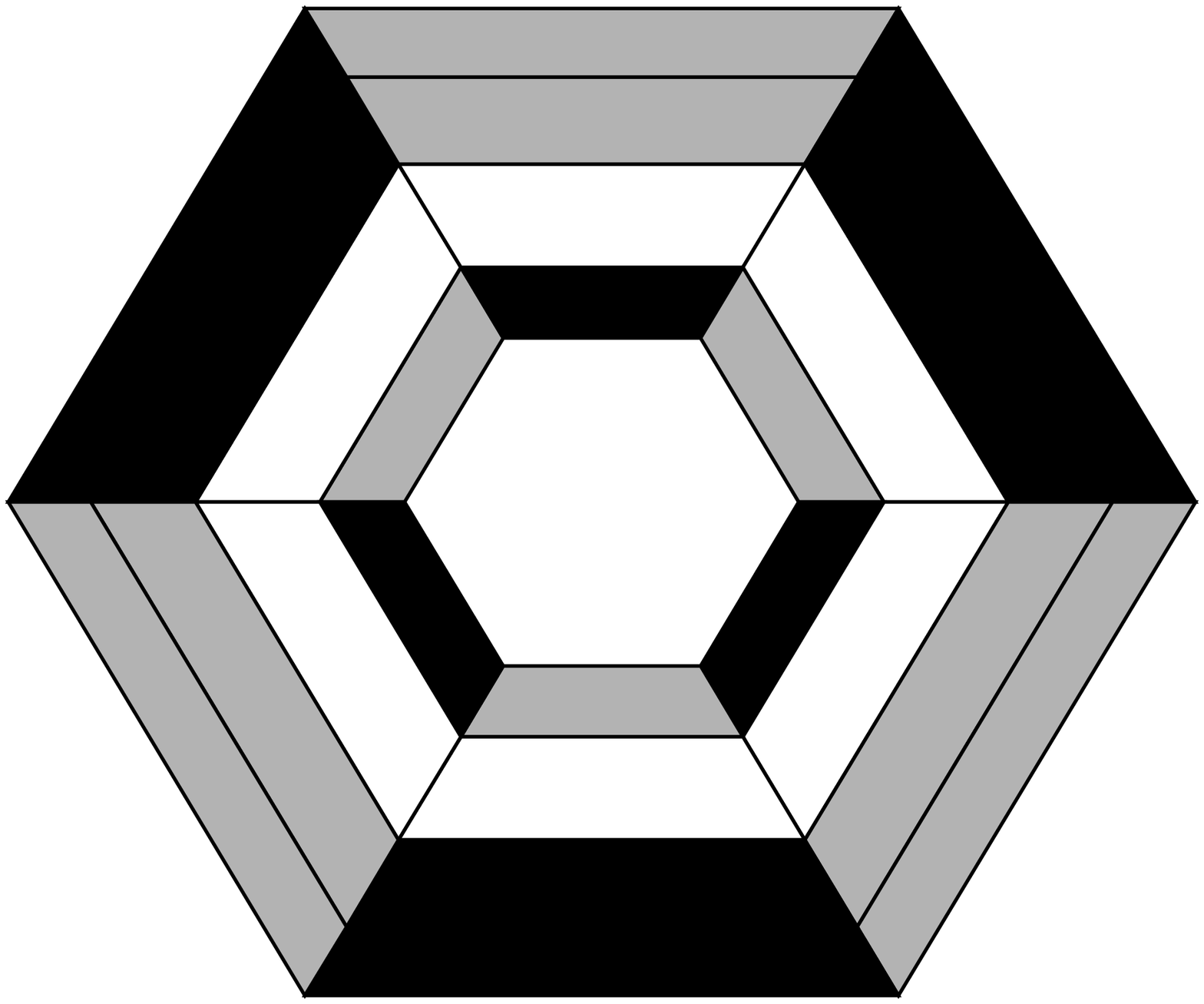}}
\medskip
\resizebox{0.48\textwidth}{!}{%
\includegraphics*[]{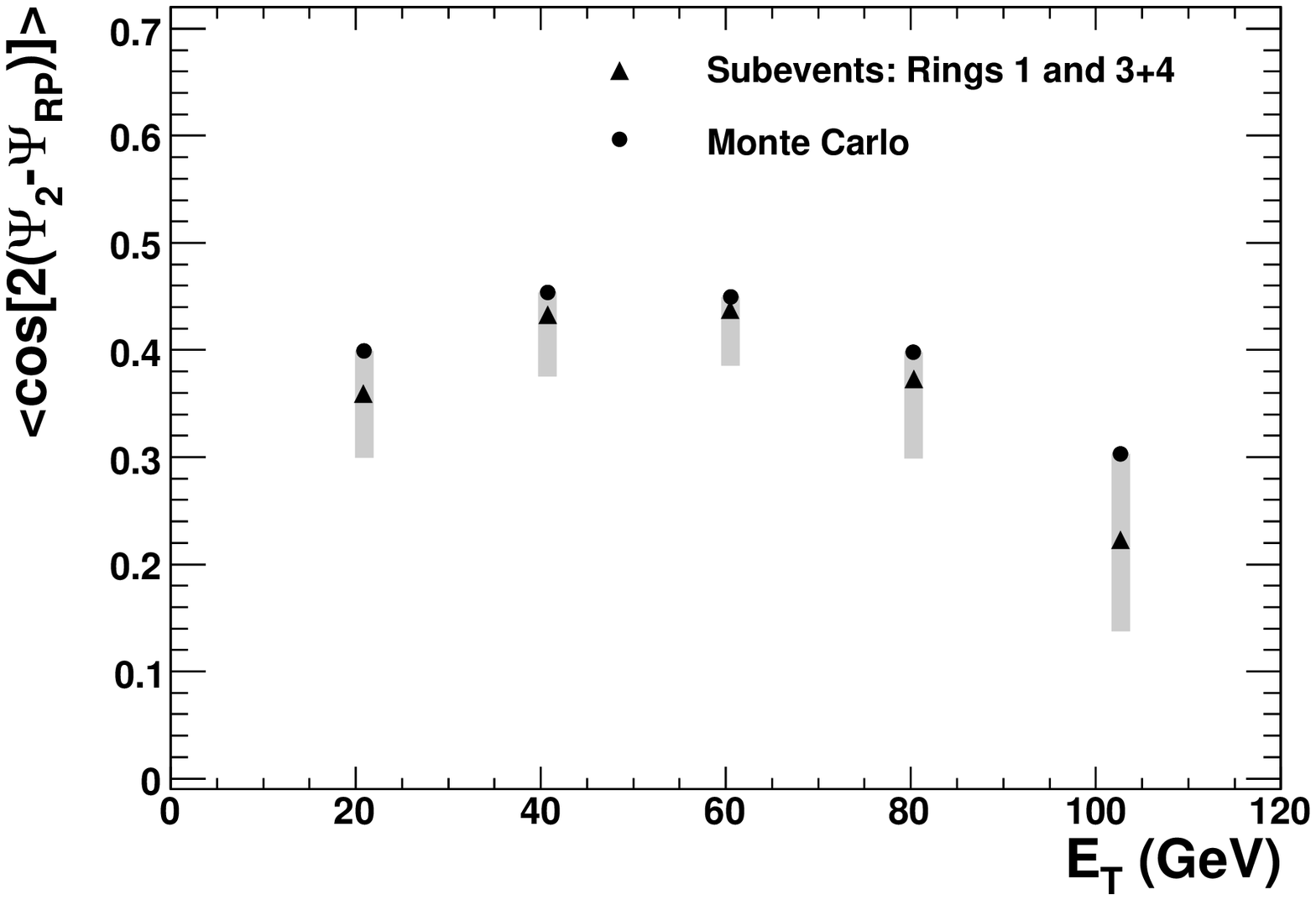}}\\
(a)\hspace{7 cm} (b)\\
\caption{(a) Sketch of the geometrical configuration used for sub-event 
definition. (b) Second harmonic event plane resolution with from Monte Carlo 
simulations and sub-events technique.}
\label{fig:evplres}
\end{figure}

The event plane resolution (expressed as 
$< \cos \left[ 2 \left ( \Psi_2 - \Psi_{RP} \right) \right] >$ )
has been estimated in two independent ways.
The results are shown in fig.~\ref{fig:evplres}b.
The first technique is based on Monte Carlo simulations of the detector 
response taking as input the value of $v_2$ measured by the 
calorimeter~\cite{NA50flow}.
The sextant to sextant fluctuations are tuned to reproduce the experimentally 
observed  distribution of the quantity:
\begin{equation}
b_3=\frac{\sum_{k=1}^{6} \sin \left[ 3 (2k-1) \frac{\pi}{6} \right] E_{Tk}}
{\pi \mathrm{sinc}\frac{3\pi}{6}}
\end{equation}
with $\mathrm{sinc}~x=(\sin x )/x$.
For symmetry reasons, $b_3$ is sensitive only to statistical fluctuations 
and not to azimuthal anisotropies~\cite{NA50flow}.
The gray band represents the systematic error coming from the 
systematic uncertainty on $v_2$ due to the presence of non-flow correlations.
The second technique makes use of the ring segmentation of the calorimeter to
define two sub-events. The resolution is extracted from the angular 
correlation between the event plane angles of the two sub-events. 
A scheme of the sub-event definition is shown in fig.~\ref{fig:evplres}a: 
ring 2 has been removed from the analysis in order to have a 
rapidity gap limiting non-flow correlations between the two sub-events, while
the alternate pattern of rings and sextants is dictated by the need of having 
the two sub-events equally populated.
The energy collected by the excluded ring 2 is then accounted for when
using the formulas from~\cite{PoskVol} to extrapolate the measured resolution 
of the sub-event plane to the resolution of the event plane of the full 
calorimeter.
As it can be seen in fig.~\ref{fig:evplres}b the resolutions extracted
with the two methods agree within the systematic uncertainties, so 
the Monte Carlo estimation together with its systematic error bar is used
to calculate the \jpsi\ elliptic anisotropy.

\section{Analysis and results}

The \jpsi\ azimuthal distribution relative to the
reaction plane is extracted using two different analysis schemes.
The first one extracts the number of \jpsi's in 
different intervals of azimuthal angle relative to the event plane.
The second one computes the Fourier
coefficient $v_2$ from the \jpsi\ azimuthal distribution.
For each analysis, specific methods are implemented in order to separate the
\jpsi\ signal from the other dimuon sources under the \jpsi\ mass 
peak\footnote{About 90\% of the reconstructed opposite-sign dimuons in the 
mass range 
$2.9<M<3.3$ GeV/c$^2$ and in the kinematic domain defined above originate 
from \jpsi\ decays.}.

\subsection{Number of \jpsi's in bins of azimuthal angle}

Two different analysis methods have been developed to extract the 
number of \jpsi's 
in two wide bins of azimuthal angle relative to the event plane
($\Delta\Phi_2=\Phi_{dimu}-\Psi_2$ where $\Phi_{dimu}$ is the 
dimuon azimuthal angle and $\Psi_2$ the second harmonic event plane 
from eq.~\ref{eq:evpl}).

\begin{figure}[bt!]
\centering
\resizebox{0.4\textwidth}{!}
{\includegraphics*[]{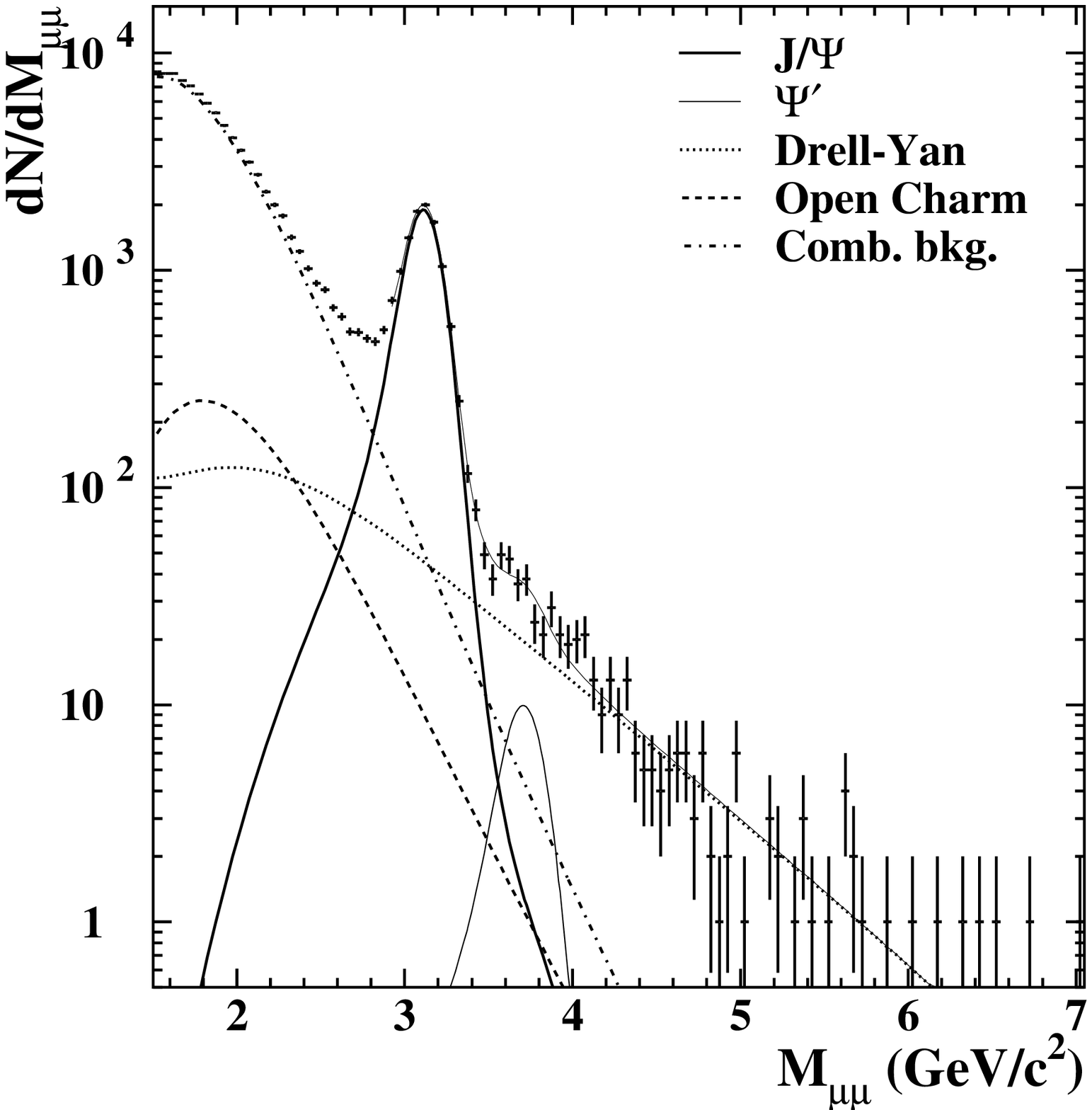}}
\resizebox{0.4\textwidth}{!}
{\includegraphics*[]{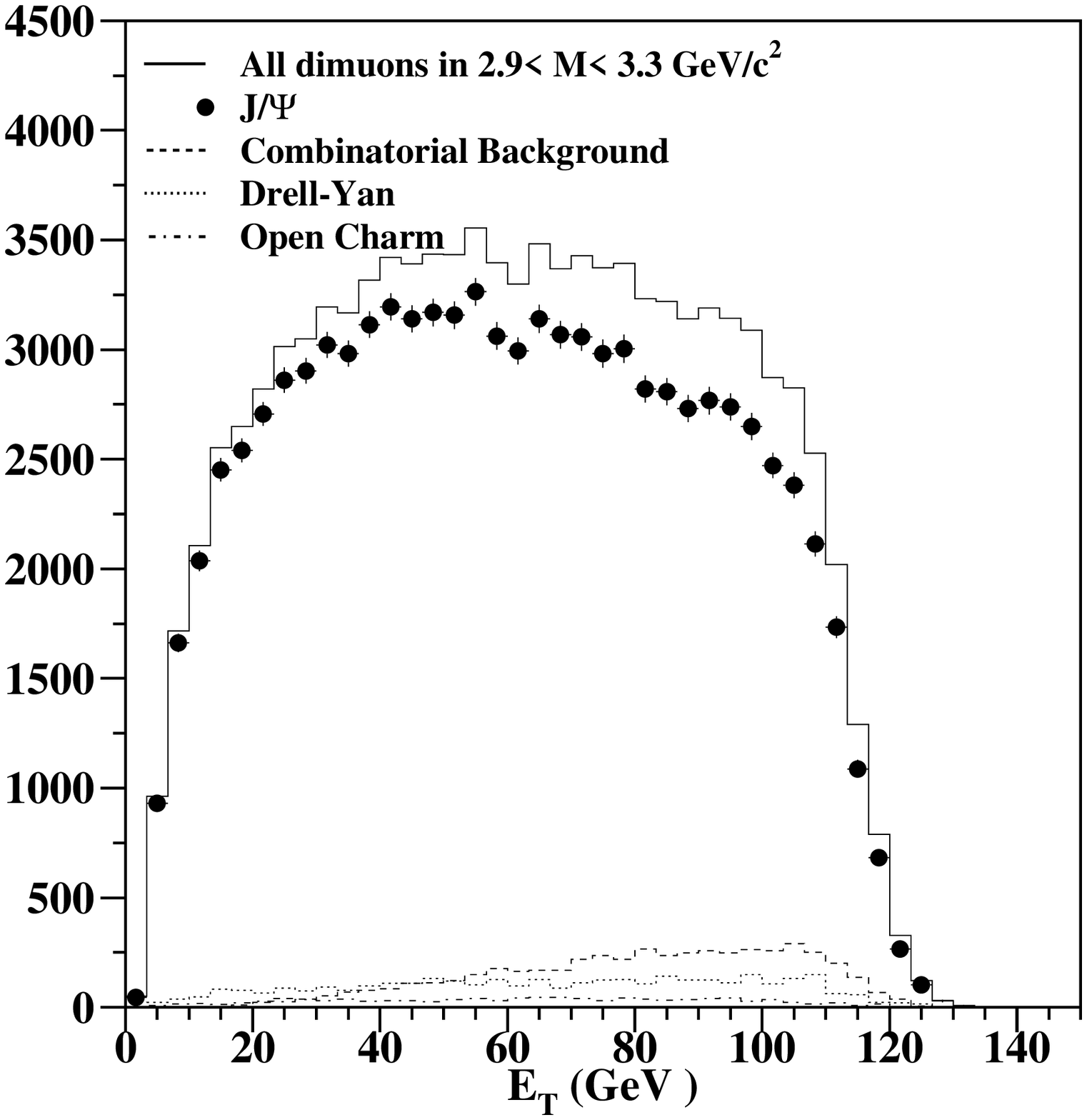}}
\caption{Left: Example of fit to the $\mu^+\mu^-$  mass spectrum. 
Right: \et\ spectra (not corrected for centrality-dependent 
inefficiencies) of $\mu^+\mu^-$ in \massrange. } 
\label{fig:spectra}
\end{figure}

The first method consists in fitting the mass spectra of opposite-sign 
dimuon sub-samples in bins of centrality (\et) and dimuon
azimuthal angle relative to the measured event plane ($\Delta\Phi_2$).
The mass spectra above 2.5~GeV/$c^2$ 
(see fig.~\ref{fig:spectra}-left) are fitted to the four signal 
contributions (namely \jpsi, \psip, Drell-Yan and open charm) with 
shapes determined from detailed Monte Carlo simulations of the NA50 apparatus. 
The combinatorial background is evaluated from the like 
sign pairs (see~\cite{Ale05} for details).
This analysis method is limited by the low statistics of high-mass
Drell-Yan dimuons which are crucial to fix the Drell-Yan contribution
in the fitting procedure.
Hence, it is not possible to divide the full sample of dimuon
events in more than 10 bins (5 centrality $\times$ 2 $\Delta \Phi_2$ intervals)

The second method consists in building, for each  $\Delta\Phi_2$ bin, 
the \et\ spectrum of all the $\mu^+\mu^-$ in the mass range 
\massrange\ and then subtracting the 
spectra of the different sources of background.
The \et\ spectra of the various dimuon contributions in \massrange\ are shown 
in fig.~\ref{fig:spectra}-right.
The combinatorial background is extracted from like-sign muon pairs.
The DY spectrum is estimated from $\mu^+\mu^-$ in the mass range 
4.2$<$M$<$7.0 GeV/c$^2$ and rescaled, via Monte Carlo simulations, 
to \massrange. 
The \ddbar\ yield is estimated from opposite sign dimuons in  
2.1$<$M$<$2.7 GeV/c$^2$ after combinatorial background and 
DY subtraction and rescaled to the \jpsi\ mass range.
The dimuons from \psip\ decay in \massrange\ are negligible. 
The underlying assumption is that the \et\ spectra of the involved physical 
processes do not depend on the invariant mass range used for their 
determination within the range under study.
This ``counting'' method allows for a larger number of \et\ bins, thus 
providing a better insight on the centrality dependence of the possible 
anisotropy.
It should be noted that no correction for centrality-dependent inefficiencies 
is applied to the \et\ spectra because in the following analyses
only relative numbers of \jpsi's in different $\Delta\Phi_2$ bins
are considered.

\begin{figure}[t!]
\centering
\resizebox{0.48\textwidth}{!}
{\includegraphics*[]{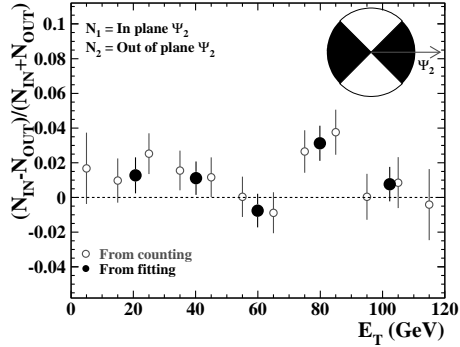}}
\caption{
Anisotropy vs. \et\ as extracted from the number of \jpsi's in 
$\Delta \varphi_n$ bins centered in-plane and out-of-plane (right).
Error bars represent statistical errors on the measured points.}
\label{fig:aniset}
\end{figure}

The presence of an azimuthal dependent \jpsi\ absorption is expected to
result in a different number of particles emitted parallel (in-plane)
and orthogonal (out-of-plane) to the reaction plane as a consequence of
the geometrical shape of the overlap region of the colliding nuclei.
The elliptic anisotropy is therefore quantified starting from
the numbers $N_{IN}$ and $N_{OUT}$ of \jpsi's observed in two cones 
with an opening angle of 90$^\circ$ centered 
respectively at $\Delta\Phi_2$=0$^\circ$ (in-plane) and at 90$^\circ$ 
(out-of-plane), see fig.~\ref{fig:aniset}.
So:
\begin{eqnarray}
N_{IN}=\int_{-\pi/4}^{\pi/4}\frac{dN}{d(\Delta\Phi_2)}d(\Delta\Phi_2)+
\int_{3\pi/4}^{5\pi/4}\frac{dN}{d(\Delta\Phi_2)}d(\Delta\Phi_2)\\
N_{OUT}=\int_{\pi/4}^{3\pi/4}\frac{dN}{d(\Delta\Phi_2)}d(\Delta\Phi_2)+
\int_{5\pi/4}^{7\pi/4}\frac{dN}{d(\Delta\Phi_2)}d(\Delta\Phi_2)
\end{eqnarray}

The anisotropy is then quantified as the ratio 
$(N_{IN}-N_{OUT})/(N_{IN}+N_{OUT})$. A positive anisotropy comes from
a larger number of \jpsi's observed in plane than out-of-plane.
If only a second (elliptic) harmonic is present,
i.e. ${dN}/{d\Phi_{dimu}} \propto 1+2 v_2 \cos[2(\Phi_{dimu}-\Psi_{RP})]$,
then:
\begin{equation}
\frac{N_{IN}-N_{OUT}}{N_{IN}+N_{OUT}}=\frac{4}{\pi}v_2
\end{equation}

It should be noted that the resolution of the event plane has not been 
taken into account in the calculation of the anisotropy from
$N_{IN}$ and $N_{OUT}$ and therefore the comparison with the values
of the Fourier coefficient $v_2$ reported in the next section
is not straightforward.
The results for the elliptic anisotropy as a function of
\et\ are shown in fig.~\ref{fig:aniset}.
The two analyses agree in indicating on average a small excess of \jpsi's 
emitted in-plane (positive anisotropy).
The largest signal is observed in the centrality bin 70$<$\et $<$ 90 GeV.

\subsection{Fourier coefficient $v_2$ of \jpsi's}

The second coefficient of the Fourier expansion is given by:
\mbox{$v_2=<\cos[2(\Phi_{dimu}-\Psi_{RP})]>$}
where the average is performed over events in a given centrality (or \pt)
bin.
Since the reaction plane ($\Psi_{RP}$) is unknown, the event plane (calculated 
from neutral transverse energy anisotropy) has to be used instead, obtaining
\mbox{$v^\prime_2=<\cos[2(\Phi_{dimu}-\Psi_{2})]>$}.
The quantity $v^\prime_2$ should then be corrected for the
event plane resolution~\cite{PoskVol}, obtaining
\mbox{$v_2=v^\prime_2/<\cos[2(\Psi_2-\Psi_{RP})]>$}.

Two different analysis methods have been used to subtract the background 
and extract the values of \jpsi\ elliptic anisotropy $v_2$.
The first estimation is obtained from the average of the 
\mbox{$\cos[2(\Phi_{dimu}-\Psi_{2})]$} distributions 
of $\mu^+\mu^-$ in the mass range \massrange\ after subtracting the background 
contributions with the same ``counting'' procedure described above. 
The \mbox{$\cos[2(\Phi_{dimu}-\Psi_{2})]$} distribution of 
combinatorial background is estimated from like-sign muon pairs, while
the ones of DY and \ddbar\ are extracted from different $\mu^+\mu^-$ mass 
intervals and rescaled to the \jpsi\ mass range
under the assumption that they do not depend on the invariant 
mass range considered.

A second evaluation of $v_2$ in the 5 \et\ bins has been 
obtained from the number of \jpsi's extracted with the ``counting'' method 
in 8 bins of azimuthal angle relative to the event plane.
The coefficient $v^\prime_2$ is obtained by fitting the resulting number 
of \jpsi's in bins of $\Delta\Phi_2$ with the function
\begin{equation}
N^{J/\psi}(\Delta \Phi_2)=K 
\left[1+2 \cdot v^\prime_2 \cos(2\Delta \Phi_2)\right]
\label{eq:fitv2}
\end{equation}
where $\Delta \Phi_2=\Phi_{dimu}-\Psi_2$ and the free parameters of the fit are
$K$ and  $v^\prime_2$. 
Afterward, $v_2$ is obtained by applying to $v^\prime_2$ the correction 
factor for the event plane resolution.

\begin{figure}[tb!]
\centering
\resizebox{0.48\textwidth}{!}
{\includegraphics*[]{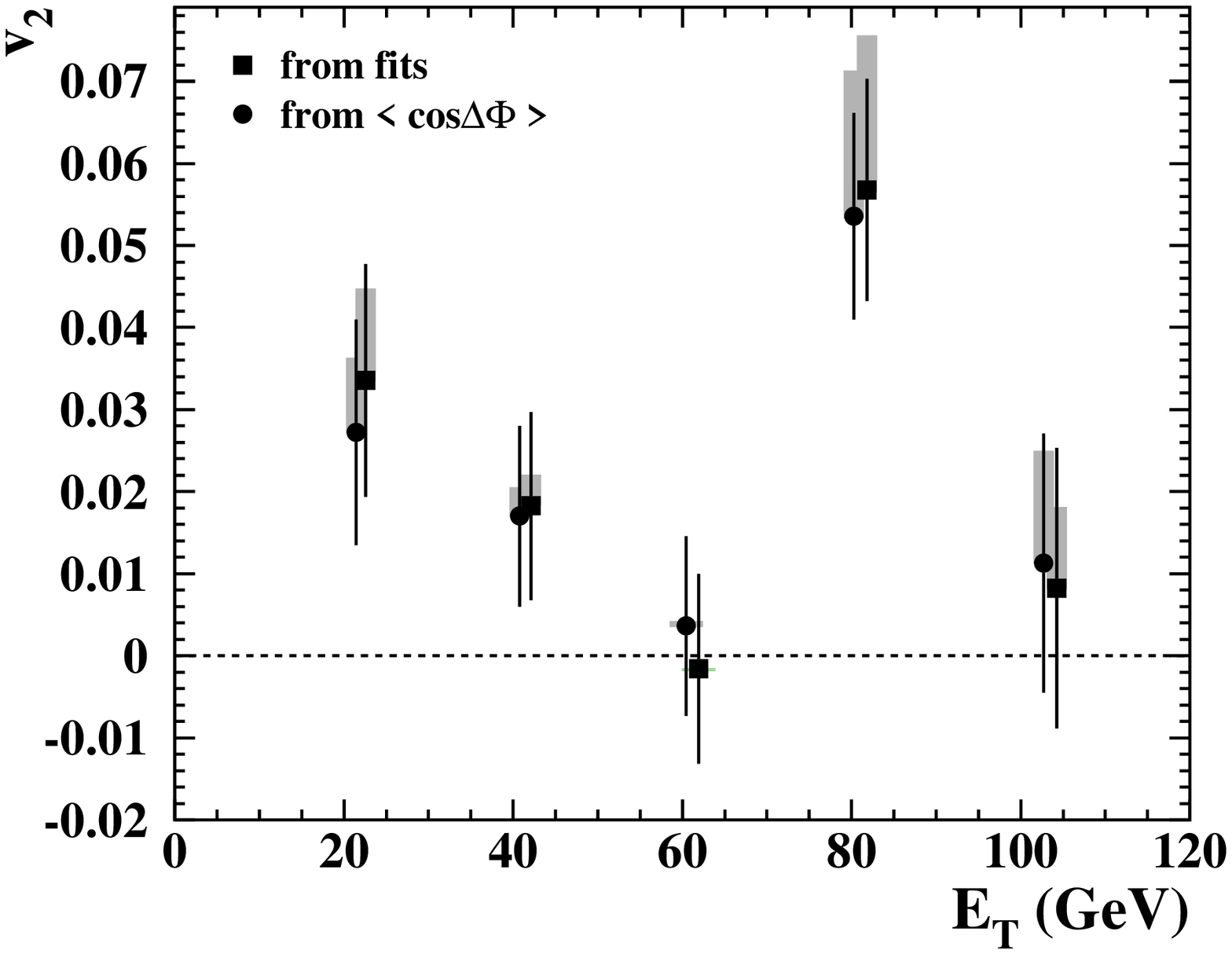}
}
\resizebox{0.48\textwidth}{!}
{\includegraphics*[]{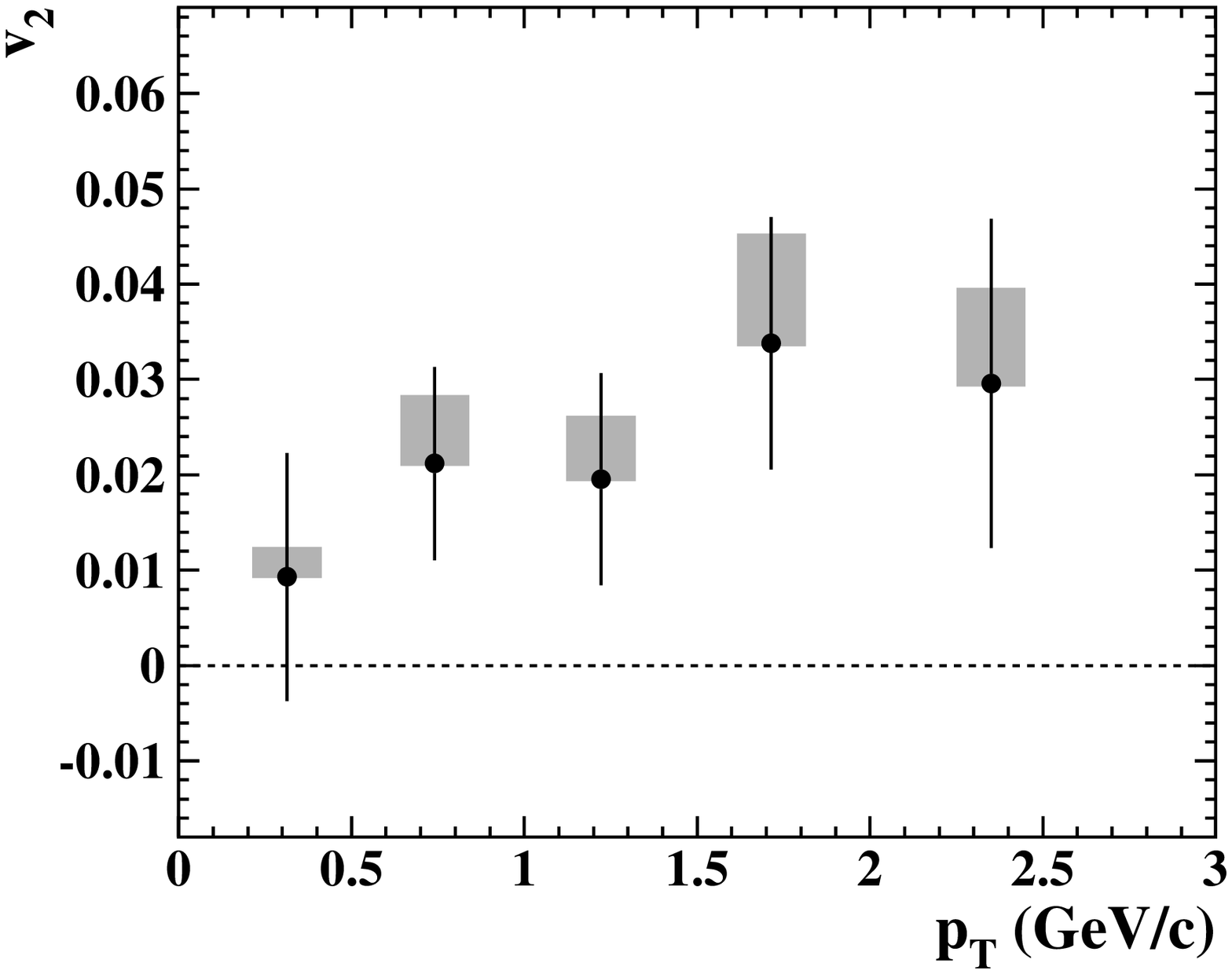}}
\caption{Fourier coefficient $v_2$ 
vs. \et\ (left) and \jpsi\ \pt (right).
Error bars represent statistical errors, 
the gray band is the systematic error coming from
the estimation of the event plane resolution.}
\label{fig:v2}
\end{figure}

The results for $v_2$ vs. \et\ are shown in fig.~\ref{fig:v2}-left. 
The two analysis methods are in remarkable agreement 
and show positive values of $v_2$,
confirming the excess of \jpsi's exiting in-plane.
A maximum $v_2$ is observed for the bin 70$<$\et$<$90 GeV, corresponding to
$N_{part}\approx270$ and $\langle$b$\rangle$=4.8 fm.
The error bars represent the statistical error on the measurement of the \jpsi\
anisotropy, while the gray bands are the systematic errors coming from the 
uncertainty on the estimation of the event plane resolution.
The analysis based on the \mbox{$\cos[2(\Phi_{dimu}-\Psi_{2})]$} 
spectra has been applied also in bins of \jpsi\ transverse momentum.
The obtained results for $v_2$ as a function of \pt\ (centrality integrated) 
are shown in~\ref{fig:v2}-right.
The \jpsi\ $v_2$ shows an increasing trend with increasing \pt.

\section{Conclusions}

\jpsi\ elliptic anisotropy relative to the reaction plane 
has been measured by NA50 from a data sample of 100000 \jpsi's produced
in Pb-Pb collisions at 158 GeV/nucleon ($\sqrt{s}=17.2$ GeV).
The anisotropy has been quantified both from the normalized difference
between the number of \jpsi's emitted in plane and out-of-plane and
from the Fourier coefficient ($v_2$) which describes an elliptic anisotropy.
These quantities have been measured as a function of collision centrality 
(defined by the neutral transverse energy \et\ produced in the collision) 
and as a function of \jpsi\ transverse momentum.
A positive $v_2$ is measured: more \jpsi's are observed in-plane than 
out-of-plane. The largest anisotropy is observed
in the centrality bin with $N_{part}\approx270$ and $\langle$b$\rangle$=4.8 fm.
The elliptic anisotropy is observed to increase with increasing \jpsi\ \pt.

\end{document}